\documentclass[aps,pra,preprint,showpacs,groupedaddress]{revtex4}
\usepackage{graphicx}
\usepackage{amsmath}
\usepackage{longtable}

\begin{document}

\title{Properties of the triplet metastable states of the alkaline-earth atoms.  }
\author{J.Mitroy}  
\author{M.W.J.Bromley}  
\altaffiliation{now at Department of Physics, Kansas State University,
Manhattan, KS 66506 USA}
\email{bromley@phys.ksu.edu}
\affiliation{Faculty of TIE, Charles Darwin University, 
Darwin NT 0909, Australia}

\date{\today}

\begin{abstract}

The static and dynamic properties of the alkaline-earth atoms 
in their metastable state are computed in a configuration 
interaction approach with a semi-empirical model potential 
for the core.  Among the properties determined are the
scalar and tensor polarizabilities, the quadrupole moment,
some of the oscillator strengths and the dispersion coefficients
of the van der Waals interaction.   A simple method for 
including the effect of the core on the dispersion parameters 
is described.  

\end{abstract}


\maketitle

\vspace{1cm}
\section{Introduction} 

The low-lying triplet metastable states of alkaline-earth 
atoms have been generating increasing interest in the area 
of cold-atom physics for a number of reasons.  One application 
is to use the $^1S^e_0$ $\rightarrow$ $^3P^o_1$ transition 
in calcium as a new optical frequency standard \cite{wilpers02a}.  
The use of the $^1S^e_0$ $\rightarrow$ $^3P^o_0$ transitions 
for fermionic $^{87}$Sr stored in an optical lattice is expected 
to further result in an improved standard \cite{katori03a}.  
Another possible application is in the formation of  
Bose-Einstein condensates (BEC) consisting of 
alkaline-earth atoms \cite{derevianko01a,derevianko03a} in
their metastable triplet states.   The stability, size and
excitation modes of BECs depends on the sign (and 
magnitude) of the scattering length, and the scattering 
length depends sensitively on the precise values of the dispersion 
constants \cite{tiesinga02a,timmermans00a}.   

Taken in conjunction, the desirability of obtaining precise 
values of the static and dynamic properties of the low-lying 
$nsnp$  $^3P^o$ metastable state has greatly increased in
importance.   In this article, properties of these states 
are computed from valence electron configuration interaction 
calculations that use  
a semi-empirical model potential to describe the core-valence 
interaction \cite{mitroy88d,mitroy93a,bromley02b,mitroy03e,mitroy03f}.  
Among the data computed are the oscillator strengths for some
of the low-lying transitions, the scalar and tensor polarizabilities,
the quadrupole moments, and the dispersion coefficients for
the van der Waals interaction between two atoms.    

\section{Results of the calculations} 

\subsection{Methodology} 

The properties of these states are computed using configuration 
interaction (CI) calculations that treat the correlations 
between the valence particles in an ab-initio manner while using 
a semi-empirical model potential to describe the core-valence 
interaction \cite{mitroy88d,mitroy93a,bromley02b,mitroy03e,mitroy03f}.  
The details of this calculation are very similar to those 
reported in \cite{bromley02a,bromley02b,mitroy03f} apart from some
minor changes in the cutoff parameters and the use of an orbital
basis of larger dimension.  The polarization potentials
were initially defined by tuning the potential to reproduce
the $ns$, $np$, $nd$ and $nf$ binding energies of the respective
singly ionized atom.
The Hamiltonian was then diagonalized in a basis consisting
of all the two-electron basis states that could be formed
from a set of about 140-150 single particle orbitals.  The basis set
contained orbitals up to and including $\ell \le 8$ and the 
two-electron basis dimensions ranged from 1000 to 4000.  For all
practical purposes the basis for the two-valence electrons
can be regarded as saturated.  The initial binding energies 
obtained by this procedure were not in perfect agreement with 
experiment with discrepancies for the ground and excited state 
energies of the order of 0.1-2.0$\%$ (refer to
\cite{bromley02a,bromley02b} to get an indication of the
accuracy).  Some further tuning of the cutoff parameters
was done to improve the accuracy of the energy differences
which directly impact on the accuracy of expectation values.
Expectation values for multipole operators were computed
with a modified operator that allowed for polarization 
corrections \cite{hameed68a,hameed72a,mitroy03f}.   

The model potential is quite realistic since the direct and 
exchange interactions with the core were computed without 
approximation from a 
Hartree-Fock (HF) wave function, only the core polarization
potential was described with a model potential.  The resulting
polarizabilities, and dispersion parameters for homo-nuclear 
pairs of atoms were generally within 0.1$\%$ of the best variational 
calculations for Li or Be, and for heavier atoms they were 
generally within 1-2$\%$ of results coming from large  
fully relativistic calculations
combining configuration interaction and many-body perturbation
theory techniques \cite{mitroy03f}.  

The most likely source of error in the present calculations
for the heavier species, Ca and Sr, is the neglect of relativistic 
effects.  However, the use of a polarization potential tuned to 
the experimental binding energy will implicitly take into
account the influence of relativistic effects upon the core
electron distribution.  Further, Greene and Aymar have shown
that the spin-orbit interaction does not have major effect on
the structure of the alkaline-earth wave functions \cite{greene91a}.   

\subsection{Energy levels} 

The energy levels of the present calculations are given in Table 
\ref{tab1} and compared with experiment.  The  polarization 
cut-off parameters were fine-tuned to reproduce the experimental
binding energy of the lowest states of each symmetry.
In the case for states with $L > 0$ the parameters were
tuned to reproduce the center-of-gravity of the spin-orbit
triplets.  The spin-orbit splitting of the triplet states
is largest for strontium and its magnitude is about 0.001 
Hartree.   

The agreement between the theoretical and experimental energy
levels is sufficiently close to discount the possibility that 
energy level considerations might make a significant contribution
to the uncertainty in the oscillator strengths and polarizabilities. 

\begingroup
\squeezetable 
\begin{table}[th]
\caption[]{ \label{tab1}
Theoretical and experimental energy levels (in Hartree) of some
of the low-lying metastable states of the alkaline-earth atoms.
The energies are given relative to the energy of the
doubly ionized core.  The experimental energies for the triplet 
states are averages with the usual $(2J+1)$ weighting factors. 
The experimental data are taken
from \cite{nistasd2,moore71c}.  The $md$ level is the $3d$ 
level for Be, Mg and Ca while for Sr it is the $4d$ level.  }  
\begin{ruledtabular}
\begin{tabular}{lcccccccc}
Systems &  \multicolumn{2}{c}{Be} & \multicolumn{2}{c}{Mg} &
          \multicolumn{2}{c}{Ca}  &  \multicolumn{2}{c}{Sr}  
\\ \cline{2-3} \cline{4-5} \cline{6-7} \cline{8-9}
Level & Theory  & Exp.  &  Theory  & Exp. &  Theory  & Exp. & Theory  & Exp.  \\
\hline
$ns^2$ $^1S^e$ &  -1.011842  & -1.011850 &  -0.833533   &  -0.833530   &   
      -0.660944   & -0.660932 & -0.614598  &   -0.614602 \\
$nsnp$ $^3P^o$  &  -0.911710  &  -0.911701 &  -0.733378     & -0.733788    & 
    -0.591387    & -0.591388 &  -0.547611  &   -0.547612 \\
$ns(n+1)s$ $^3S^e$  & -0.774561    & -0.774552 &  -0.645827   &  -0.645821   &  
     -0.517230  &  -0.517228   &  -0.482289  &    -0.482292 \\
$np^2$ $^3P^e$  &  -0.739862    & -0.739855 &  -0.569906  & -0.569929 &   
     -0.485477  & -0.485478 & -0.452720   &     -0.452717 \\
$nsmd$ $^3D^e$  & -0.729118  & -0.729113  &  -0.615041    &  -0.615022 &  
     -0.568193  & -0.568180 & -0.531359   &   -0.531367 \\
\end{tabular}
\end{ruledtabular}
\end{table}
\endgroup

\subsection{Quadrupole moments} 

The quadrupole moment of the $^3P^o_2$ state is a static property of 
the state.  An exact knowledge of its value is important since the
quadrupole-quadrupole interaction has a big impact of the cold 
collision physics in metastable alkaline-earth metal atoms 
\cite{derevianko03a,kokoouline03a}.  Defining the $LS$ coupled 
reduced matrix element as    
\begin{equation}
Q(L) =  \langle \Psi(^3P^o) \parallel \sum_i r_i^2 \mathbf{C}^2(\mathbf{r}_i) \parallel \Psi(^3P^o) \rangle   \ ,  
\label{QLmomt}
\end{equation}
the quadrupole moment for a triplet state is usually defined  
as the moment of $^3P^0_J$ state with $M_J = J$; In the expression 
above ${\bf C}^2(\mathbf{\hat r})$ is the spherical tensor of 
rank 2. 
\begin{equation}
Q =  \langle \Psi(^3P^o_J); M_J=J |  
         \sum_i r_i^2 C^2_0(\mathbf{r}_i) | \Psi(^3P^o_J);M_J=J \rangle   \ ,  
\label{Qmomt}
\end{equation}
This can be written \cite{santra03a}    
\begin{eqnarray}
Q(LJ) &=& \sqrt{\frac{4J(2J-1)}{(J+1)(2J+1)(2J+3)}} (2J+1) (-1)^{2+S+L+J} 
\begin{Bmatrix} S & L & J \\ 2 & J & L \end{Bmatrix} 
            Q(L)  \ ,  
\label{QJmomt}
\end{eqnarray}
where the Wigner-Eckart theorem has been used twice to collapse
the angular factors.  
The quadrupole moment for a $^3P$ level is often given
for the $J = 2$ state.  The quadrupole moment $Q(LJ)$ for the 
$J =2$ state is equal to $2\langle Q_{zz} \rangle$.
  
Some older calculations of $\langle Q_{zz} \rangle$ exist 
\cite{ceraulo91a,sundholm93a}.  The finite element
Multi Configuration Hartree-Fock (MCHF) calculation of Sundholm
and Olsen for Be gave 4.53 au which is in excellent agreement
with the present value of 4.54 au.  The CI calculations of
Ceraulo and Berry \cite{ceraulo91a} consistently underestimated 
the present quadrupole moments (e.g. 7.944 au for Mg) and are 
not listed in Table \ref{properties}.

The quadrupole moments are compared with the recent calculations 
by other groups in Table \ref{properties}. The CI+MBPT calculation
\cite{derevianko03a} is a fully relativistic calculation with 
the post-HF interactions between the valence electrons and the core 
treated with perturbation theory while the interaction between 
the two valence electrons are teated with the CI ansatz.  

The calculation of Santra and Greene \cite{santra04a} (SG-CI) treated 
the two active electrons within a CI framework while using a model 
potential to represent the core-valence interaction.  The model 
potential did include a spin-orbit interaction.  One limitation with 
the SG-CI calculation is that it does not include the di-electronic 
part of the polarization potential.    

The noticeable feature of Table \ref{properties} is that all three 
calculations agree with other with a total variation of less than
2$\%$.  The present results generally lie closer to the CI+MBPT 
calculation than the SG-CI calculation.  The high level of agreement
between three completely independent calculations suggests that the 
uncertainty ascribed by Derevianko {\em et al} to their quadrupole
moment was too big by a factor of 2. 

\subsection{Oscillator strengths of low-lying transitions} 

The oscillator strengths for the transitions to the lowest
lying  $^3S^e$, $^3P^e$ and $^3D^e$ states are given in 
Table \ref{properties}.  The absorption oscillator strength 
from state $\psi_0$ is calculated according to the
identity

\begin{equation}
f_{0n} =  \frac {2 |\langle \psi_0;L_0 S \parallel \sum{_i} \  r_i 
{\bf C}^{1}({\bf \hat{r}}_i) \parallel \psi_{n};L_n S\rangle|^2 \epsilon_{0n}}
{3(2L_0+1)}  \ .
\label{fvaldef}
\end{equation}
The oscillator strengths for the Be triplet transitions are 
probably as accurate as any that have previously been published. 
The basis for the valence electrons is effectively saturated
and the semi-empirical approach to core polarization is capable 
of high accuracy \cite{mitroy03f}.     
For example, the present methodology reproduces the dipole and 
quadrupole polarizability of Be given by a close to exact calculation
\cite{komasa02a} to an accuracy of 0.2$\%$.  The present 
oscillator strengths agree very well with the experimental 
values given in Table \ref{properties}.   
Not shown in the Table are the $^3P^o$ $\rightarrow$ $^3P^e$ 
oscillator strengths of CI calculation of Weiss \cite{weiss95a} 
and the MCHF calculation of Jonsson 
{\em et al} \cite{jonsson99a}.  Both of these calculations 
were very large and incorporated both core and valence 
excitations.  The Weiss $f$-value was 0.447, while the 
Jonsson {\em et al} result was 0.4463.  These could 
hardly be any closer  to the present value of 0.4467.
 
The present oscillator strength for the transition to the $^3S^e$ 
state in Mg, namely 0.138 is in excellent agreement with that obtained
from the low uncertainty experiment of Andra {\em et al} \cite{andra79a}, 
0.139$\pm$0.003.  Agreement with the large basis CI calculation of
Moccia and Spizzo (MS-CI) is also good \cite{moccia88b}.  The MS-CI 
calculation is similar to the present calculation in that 
excitations are only permitted for the valence electrons.  It does not 
allow for core-valence correlations so the present approach, which does,  
should be regarded as being more reliable.    
  
In the case of Ca, good agreement is achieved with the model potential 
calculations of Hansen {\em et al} \cite{hansen99a} for the transitions to the 
$^3S^e$ and $^3P^e$ states. A 7$\%$ discrepancy occurs for the
transition of the $4s3d$ $^3D^e$ state. The larger difference here is
expected since the $3d$ orbital does have a tendency to penetrate
into the core and therefore degrade the accuracy associated with model
potential methods.  The best ab-initio calculation is the 
MCHF calculation by Froese-Fischer and Tachiev \cite{fischer03a}.
The MCHF calculation allows for core-valence correlations and also includes 
relativistic effects using the Briet-Pauli Hamiltonian.  The MCHF oscillator
strengths listed in Table \ref{properties} are a weighted average of the 
individual lines in the multiplet.  The largest difference between the 
present and MCHF oscillator strengths is less than 4$\%$.     

The multi-channel quantum defect theory (MQDT) calculations of Werji 
{\em et al} \cite{werji92a} which use an $R$-matrix calculation to
determine the short-range parameters.  Their transition rate data 
was converted to oscillator strengths using experimental energy
differences and lie within 2-3 $\%$ of the present oscillator 
strengths.

The most precise 
experiment for Sr is that of Andra {\em et al} \cite{andra75a} which gave 
a lifetime of $7.89\pm0.05$ ns for the $5p^2$ $^3P^e_2$ state.  This state 
can decay to the both the $5s5p$ and $5s6p$ levels and the lifetime
was converted to an oscillator strength by neglecting the transition to
the $5s6p$ state.  This assumption is justified since the dipole 
matrix element will be small due to the $\langle 5p | 6p \rangle$ 
overlap, and the $5s6p^2$ $^3P^o$ $\rightarrow 5p^2$ $^3P^e$ energy 
difference of 0.0073 Hartree is also small.     

The comparison with the time-dependent gauge independent (TDGI) 
calculations of Merewa {\em et al} \cite{begue99a} is mainly of 
interest because these authors also give estimates of the scalar 
and tensor polarizabilities.  A quick comparison of TDGI $f$-values 
with other results in Table \ref{properties} reveals that their 
oscillator strengths do not have the same level of accuracy as
the other calculations.  The underlying atomic structure information 
entering the TDGI formalism comes from CI calculations.    

\begingroup
\squeezetable 
\begin{table*}[th]
\caption[]{  \label{properties}
Properties of the metastable $^3P^o$ levels of the alkaline-earth 
atoms and He (note, the lowest $^3P^o$ level is not metastable in
He).  The oscillator strengths to the lowest $^3S^e$, $^3P^e$ 
and $^3D^e$ states are given as $f(^3L^e)$. The scalar and tensor 
dipole polarizabilities are $\alpha_0$ and $\alpha_{2,L_0L_0}$ 
respectively. The quadrupole moment $Q$ is given for the $^3P^o_2$ 
state while the dispersion parameter $C_6$ is that for two 
$^3P^o_0$ states.  The He "Other Theory" row reports the 
results of close to exact calculations with the exception of 
$\alpha_{2,L_0L_0}$.  The present oscillator strength to the He 
$^3D^e$ state is not to a physical state, rather it is to the 
lowest energy pseudo-state.  All quantities are in atomic units 
and the numbers in brackets are the uncertainties in the last digits.  
}
\vspace{0.1cm}
\begin{ruledtabular}
\begin{tabular}{lccccccc}
Method & $f(^3S^e)$ & $f(^3P^e)$ & $f(^3D^e)$ & 
$Q$ & $\alpha_0$ & $\alpha_{2,L_0L_0}$  &  $C_6$   \\ \hline
\multicolumn{8}{c}{He} \\ 
Present & -0.1797 &       &  0.6251  & 10.264  & 46.66  & 69.62  &  5102  \\
Other Theory  & -0.1797 \cite{drake96a} &   & 0.6102 \cite{drake96a}  & 10.265 \cite{yan94a} & 46.71 \cite{yan00a} & 67.09  \cite{rerat94a} &   \\ \hline  
\multicolumn{8}{c}{Be} \\ 
Present& 0.08187 &  0.4467 &  0.2948  & 4.54  & 39.02  & 0.558  & 220.3  \\
MCHF \cite{themelis95a}&  &  &    &   &  39.33 & 0.47    \\   
TDGI \cite{merawa98a}& 0.026 &  & 0.154   &   &  36.08 & 1.04    \\   
B-spline CI \cite{chen98a}& 0.0823 & 0.453 & 0.295  &   &   &    \\   
Experiment  &  0.089(3) \cite{bromander71a} & 0.44(2) \cite{andersen71a}  & 0.29(1)\cite{bromander71a}  &  &   \\ \hline 
\multicolumn{8}{c}{Mg} \\  
Present & 0.1383 &  0.6167 &  0.6287  & 8.44  & 101.9  & -14.24  & 1004  \\ 
CI+MBPT \cite{derevianko03a} &   &    &  & 8.46(8)  &      &     &  \\  
SG-CI \cite{santra04a} &   &    &  & 8.38  &  &     & 980(30) \\  
TDGI \cite{merawa01a}& 0.136 &  & 0.625   &   &  90.7 & -19.64 &     \\   
MS-CI \cite{moccia88b} & 0.1354  & 0.6383   & 0.6336  &   &      &     & \\  
Experimental & 0.139(3) \cite{andra79a} & 0.55(4) \cite{lundin73a}  &   0.62(4) \cite{kwiatkowski80a} &    &  &   &  \\ \hline 
\multicolumn{8}{c}{Ca} \\  
Present& 0.1582 &  0.5071 &  0.08136 & 12.96  & 295.3  & -28.36  & 3363 \\
CI+MBPT \cite{derevianko03a} &   &    &  & 12.9(4)  &      &     &  \\  
SG-CI \cite{santra04a} &   &    &  & 12.7  &  &     & 3020(200) \\  
TDGI \cite{merawa01b}& 0.163 &  & 0.051   &   &  276 &  -50.0   \\   
CI+model \cite{hansen99a}& 0.1526 & 0.5030 & 0.0873   &   &   &     \\   
MCHF \cite{fischer03a}& 0.161 & 0.525 & 0.0806   &   &   &     \\   
Experimental & 0.12(2) \cite{kostin64a} &  0.522(13) \cite{smith88a} &    &   &   &  &    \\ \hline 
\multicolumn{8}{c}{Sr} \\  
Present& 0.1788 &  0.4727 &  0.08254 & 15.51  & 494.8  & -53.84  & 6074  \\
CI+MBPT \cite{derevianko03a} &   &    &  & 15.6(5)  &     &     &   \\  
SG-CI \cite{santra04a} &   &    &  & 15.4  &   &    & 5260(500) \\  
MQDT \cite{werji92a} & 0.173  &   & 0.0849  &   &   &    &  \\  
Experimental &  0.188(10) \cite{havey77a}  & 0.438(4) \cite{andra75a}  &     &   &     &   &  \\  
\end{tabular} 
\end{ruledtabular}
\end{table*}
\endgroup

\subsection{The polarizabilities} 

\subsubsection{Theoretical treatment of polarizabilities} 

This analysis is done under the premise that spin-orbit
effects are small and the radial parts of the wave functions 
are the same for the states with different $J$.   

The Stark energy shifts for the different $L_0$ levels in an 
electric field $F$ are written as \cite{angel68a}  
\begin{equation}
\Delta E = -\frac{1}{2}\alpha_{L_0M_0} F^2 \ .  
\label{starkshift}
\end{equation}
The Stark shifts for the different $M_0$ states of the $^3P^o$ level 
are different and the polarizability is written as 
\begin{equation}
\alpha_{L_0M_0} = \alpha_0 + \frac{3M_0^2-L_0(L_0+1)}{L_0(2L_0-1)}\alpha_2 \ .  
\label{alphaLM}
\end{equation}
where $\alpha_2$ is taken from the state with $M_0 = L_0$.  
The total polarizability is written in terms of both a scalar 
and tensor polarizability.  The scalar polarizability 
represents the average shift of the different $M$ levels while
the tensor polarizability gives the differential shift.    

In terms of second order perturbation theory, the energy shift 
from an electric field, $F$ pointing in the $z$-direction is  
\begin{equation}
\Delta E = \frac{1}{2} \sum_{n} 
\frac{ 2 \langle\psi_0; L_0 M_0 | \sum_i \ r_i C^1_0( {\bf{\hat r}}_i) 
| \psi_n; L_n M_n \rangle
\langle \psi_n; L_n M_n | \sum_i r_i C^1_0( {\bf{\hat r}}_i) 
| \psi_0; L_0 M_0 \rangle F^2 } {(E_0-E_n)} \ .  
\label{starkme1}
\end{equation}
The polarizability can therefore be written 
\begin{equation}
\alpha_{L_0M_0} = \sum_{n} 
\begin{pmatrix} L_0 & 1 & L_n \\ -M_0 & 0 & M_n \end{pmatrix}^2  
\frac{ 2 | \langle\psi_0; L_0 \parallel \sum_i \ r_i {\bf C}^1( {\bf{\hat r}}_i) 
\parallel \psi_n; L_n \rangle |^2 } {(E_0-E_n)}  
\label{starkme2}
\end{equation}
where the Wigner-Eckart theorem has been used to isolate the 
$M$-dependent terms.   Using the definition of the oscillator
strength, eq.~(\ref{fvaldef}) and taking the average of the energy 
shifts leads to the usual definition as a sum rule over the 
oscillator strengths.  It is   

\begin{equation}
\alpha_0 = \sum_{M_0=-L_0}^{L_0} \alpha_{L_0 M_0}/(2L_0+1) =  \sum_{n} \frac {f_{0n} } {\epsilon_{0n}^2} \ ,  
\label{alpha0L}
\end{equation}
where the sum includes both valence and core excitations and 
$\epsilon_{0n} = (E_0-E_n)$.  
The $f$-value distribution for the core was estimated using a 
semi-empirical method \cite{mitroy03f}.  In this approach one
writes 
\begin{equation}
\alpha_{core} = \sum_{i \in core} \frac {N_i} {(\epsilon_i+\Delta)^2} \ ,  
\label{alphacore}
\end{equation}
where $N_i$ is the number of electrons in a core orbital, $\epsilon_i$ 
is the Koopman energy, and $\Delta$ is an energy shift parameter
chosen so that eq.~(\ref{alphacore}) reproduces an accurate estimate
of the core polarizability determined my other, independent means.   
  
Since the $M$-dependent part of the polarizability is a tensor of rank 
2 and it is easiest to define it in terms of $\alpha_{2,L_0L_0}$.   
\begin{eqnarray}
\alpha_{2,L_0M_0}& = & \alpha_{2,L_0L_0} \times  
(-1)^{L_0-M_0} \begin{pmatrix} L_0 & 2 & L_0 \\ -M_0 & 0 & M_0 \end{pmatrix}   
\Bigg/ \begin{pmatrix} L_0 & 2 & L_0 \\ -L_0 & 0 & L_0 \end{pmatrix}  \\  
        & = & \alpha_{2,L_0L_0} \times \frac{3M_0^2-L_0(L_0+1)}{L_0(2L_0-1)} \ ,   
\label{alpha2LM}
\end{eqnarray}
where $\alpha_{2,L_0L_0}$ is    
\begin{equation}
\alpha_{2,L_0L_0} = 
\sum_n \Biggl[ \begin{pmatrix} L_0 & 1 & L_n \\ -L_0 & 0 & L_0 \end{pmatrix}^2 
-\frac{1}{3(2L_0+1)} \Biggl] 
\frac{ 2 | \langle\psi_0; L_0 \parallel \sum_i \ r_i {\bf C}^1( {\bf{\hat r}}_i) 
\parallel \psi_n; L_n \rangle |^2 } {(E_0-E_n)} \ . 
\label{alpha2LL}
\end{equation}
In terms of an $f$-value sum, this reduces to  
\begin{equation}
\alpha_{2,L_0L_0} = -\biggl( \sum_{n,L_n=0} \frac {f_{0n} } {\epsilon_{0n}^2}   
           - \frac{1}{2} \sum_{n,L_n=1} \frac {f_{0n} } {\epsilon_{0n}^2}    
           + \frac{1}{10} \sum_{n,L_n=2} \frac {f_{0n} } {\epsilon_{0n}^2} \biggr) \ .  
\label{alpha2L}
\end{equation}
The core does not make a contribution to the tensor polarizability
since it has an equal impact on all the different $M$-levels.   

The development above is for $LS$ coupled states, but it
is common to give the tensor polarizability for $LSJ$ states.  
These can be related to the $LS$ states by geometric factors
arising from the application of Racah algebra.  The polarizability
can be expanded    
\begin{equation}
\alpha_{J_0M_0} = \alpha_0 + \frac{3M_0^2-J_0(J_0+1)}{J_0(J_0-1)}\alpha_{2,J_0J_0} \ .  
\label{alphaJM}
\end{equation}
where $\alpha_{2,J_0J_0}$ is the tensor polarizability of the state with 
$M_0 = J_0$.  The scalar polarizability for the different $J$ levels are 
the same and equal to the scalar polarizability in the $L$ representation. 
The tensor polarizability between the $L$ and $J$ representations
can be related by  
\begin{equation}
\alpha_{2,J_0J_0} =  \alpha_{2,L_0L_0}  
(2J_0+1) (-1)^{S+L_0+J_0+2} 
\begin{Bmatrix} S & L_0 & J_0 \\ 2 & J_0 & L_0 \end{Bmatrix} 
\frac{ \begin{pmatrix} J_0 & 2 & J_0 \\ -J_0 & 0 & J_0 \end{pmatrix} }  
{ \begin{pmatrix} L_0 & 2 & L_0 \\ -L_0 & 0 & L_0 \end{pmatrix} } \ . 
\label{alphaJJ}
\end{equation}
When $J = 1$ this reduces to $\alpha_{2,J_0J_0} = -\alpha_{2,L_0L_0}/2$.    
(This result has been checked by converting our $LS$ coupled 
$f$-values into $LSJ$ coupled values and then using
the standard expression in terms of the  
$| \langle J_0 \parallel {\mathbf r} \parallel J_n \rangle |^2$ 
matrix elements \cite{angel68a}.)    

\subsubsection{Results of calculations} 

The program logic and associated numerics were initially tested 
by estimating the polarizabilities of the $1s2p$ $^3P^o$ level of
He.  The present $\alpha_0$ of 46.6 $a_0^3$ is within $0.2\%$ of the 
close to exact calculation of Yan {\em et al} \cite{yan00a}.   
Agreement with the TDGI $\alpha_{2,L_0L_0}$ of Rerat and Pouchan  
\cite{rerat94a} is not as good, but it should be noted
that the TDGI calculation calculation obtains an $\alpha_0$ of 
49.5 $a_0^3$, indicating that the Rerat-Pouchan calculation
is not quite converged.  

The present estimates of the Be polarizabilities are the most 
accurate that have been published.  The agreement with the
Themelis and Nicolaides MCHF calculation \cite{themelis95a} for 
$\alpha_0$ is reasonable, but they give an $\alpha_{2,L_0L_0}$ that 
is about 20$\%$ smaller.  This level of agreement is acceptable 
given that the MCHF calculation was much smaller, the $2s2p$
$^3P^o$ state was represented by a 3 configuration MCHF wave 
function while 14 configurations were used to represent
the excited states.  

Only a moderate level of agreement is achieved with the TDGI 
polarizabilities for Be, Mg and Ca \cite{merawa98a,merawa01a,merawa01b}.  
The static polarizabilities agree at the 10$\%$
level while the TDGI estimates of the tensor polarizability 
are up to 50$\%$ different.  The lower level of accuracy 
achieved by the TDGI calculations is consistent with the
earlier discussion concerning the accuracy of the oscillator
strengths.      

A recent measurement of the tensor polarizability for the 
$^3P^o_1$ state of Ca using an atomic polarization interferometer 
gave $2.623\pm0.015$ $\text{kHz}/(\text{kV}/\text{cm})^2$
or $10.54 \pm 0.06$ $a_0^3$ \cite{yanagimachi02a}.  The 
tensor polarizability of the $J = 1$ state is determined 
from the Ca entry in Table \ref{properties} by multiplying 
by -${\scriptstyle \frac{1}{2}}$ according to 
eq.~(\ref{alphaJJ}).  The present calculation gives 
$\alpha_{2,J_0J_0} = 14.2$ $a_0^3$ for the $^3P^o_1$ state. 
A very early estimate of the tensor polarizability for this 
state was  $12.9 \pm 3.2$ $a_0^3$ \cite{oppen70a} and
another independent experiment gave 
$\alpha_{2,J_0J_0} = 12.1 \pm 0.8$ $a_0^3$ \cite{zeiske95a}. 

The scaler polarizability of the $^3P^o$ state has not been measured
directly, but there have been measurements of the difference
between the polarizabilities of the $4s^2$ $^1S^e_0$ ground state 
and the $^3P^o_1$ ($m = 0$) ground state.
Morinaga {\em et al} \cite{morinaga96a} obtained $90.4 \pm 13.5$ 
$a_0^3$ for the difference in the polarizabilities.     
Using the polarizability of 159.4 $a_0^3$ for the Ca ground
state \cite{mitroy03f}, and the present $^3P^o_1$ ($m = 0$) 
polarizability of $295.3 - 2 \times 14.2 = 266.9$ $a_0^3$ 
gives 107.5 $a_0^3$ for the difference in the polarizability. 

The Stark frequency shift of Li and van Wijngaarden of 
$12.314 \pm 0.041$ $\text{kHz}/(\text{kV}/\text{cm})^2$.  
for the $4s^2$ $^1S^e_0$ $\rightarrow$ $^3P^o_1$ ($m = 0$) 
transition \cite{li96a} converts to a polarizability difference 
of $98.98 \pm 0.33$ $a_0^3$.    

Taken together, present estimates of $\alpha_{2,J_0J_0}$ are 
larger than experiment by about 20$\%$ while estimates of the 
$\alpha_0(4s^{2}$ $^1S^e)$ $-$ $\alpha_0(4s4p$ $^3P^o_1)$ 
polarizability difference are about 10$\%$ too large.  
Rectifying the situation in a non-relativistic calculation could 
be problematic since an improvement in $\alpha_{2,J_0,J_0}$ will
result in the theoretical polarizability difference drifting 
further away from the experimental polarizability difference.    

The obvious improvement that could eliminate this problem 
would be the inclusion of the spin-orbit interaction.  The
largest contribution to the polarizability comes from the 
transitions to the $^3D^e$ levels.  The spin-orbit splitting
leads to the excitation energies for $4s3d$ states with differing 
$J$ fluctuating by about $\pm2\%$.  Given the cancellations
that occur in the evaluation of eq.~(\ref{alpha2LL}) it is 
possible that introduction of spin-orbit splitting could lead
to a Ca tensor polarizability in better agreement with experiment.
   
There have been no measurements of the tensor polarizability for
the other alkaline-earth atoms.  This should be rectified since it
would be a very useful diagnostic with which to assess the accuracy
of the structure models of the metastable states.    

\subsubsection{Alternate treatment of core} 

It is desirable to partition the core $f$-value some into 
contributions that arise from excitations to final states
with different core+valence angular momentum, $L_T$.  
Therefore, it is possible to write symbolically 
\begin{equation}
\alpha_{core} = \sum_{L_T} \alpha_{core,L_T} \ ,  
\label{core1}
\end{equation}
where  $\alpha_{core,L_T}$ will include all the 
contributions from the different magnetic sub-levels, i.e.    
\begin{equation}
\alpha_{core,L_T} = \sum_{M_T} \alpha_{core,L_TM_T}  \ . 
\label{core2}
\end{equation}
For any of the core dipole excited magnetic sub-levels one can write 
\begin{equation}
\alpha_{core,L_TM_T} = \sum_{n} \frac { f(00:LM \rightarrow n;L_TM_T) } {\epsilon_{0n}^2} \ .  
\label{core3}
\end{equation}
The final states, $|L_TM_T\rangle$ can be expanded in terms of 
uncoupled states, e.g.       
\begin{equation}
|L_T M_T\rangle = \sum_{m M} \langle 1 m L M |L_T M_T\rangle  | 1 m L M \rangle \ ,  
\label{core4}
\end{equation}
where $1m$ refers to the angular momentum of the excited core and 
$LM$ refers to the angular momentum of the $^3P^o$ metastable
state which is acting as a spectator.  Therefore, it is possible 
to decompose the oscillator strength as   
\begin{equation}
f(0;00LM \rightarrow n;L_T M_T) = \sum_{m M} 
                           |\langle 1 m L M |L_T M_T\rangle  |L_T M_T\rangle|^2    
f(0;00LM \rightarrow n;1mLM) \ .  
\label{core5}
\end{equation}
The polarizability can also be expanded in terms of uncoupled states  
\begin{equation}
\alpha_{L_T M_T} = \sum_{m M} |\langle 1 m L M |L_T M_T\rangle  |L_T M_T\rangle|^2  
\alpha_{core,mM}  / (2L+1)  \ .   
\label{core6}
\end{equation}
The factor of $(2L+1)$ in the denominator arises due to the 
sum over spectator states.  
We now assume that the excitations for the core occur 
independently of the state of valence electrons which
act as spectators.  Therefore, the contribution to the 
polarizability is independent of $M$.  Further, the 
core initially has a net angular momentum of zero and
therefore the different magnetic sub-levels of the 
core excitations should give equal contributions to
the polarizability, hence    
\begin{equation}
\alpha_{core,mM} = \frac{ \alpha_{core} } {3}   \ .  
\label{core7}
\end{equation}
The final result is  
\begin{equation}
\alpha_{L_T M_T} = \sum_{m M} |\langle 1 m L M |L_T M_T\rangle  |L_T M_T\rangle|^2    
\frac{\alpha_{core}}{3(2L+1)} \ , 
\label{core8}
\end{equation}
which can be simplified by summing the Clebsch-Gordan coefficients to give 
\begin{equation}
\alpha_{core,L_T} = \frac{ (2L_T+1) \alpha_{core}}{ 3(2L+1) } \ .    
\label{core9}
\end{equation}

When particular values are substituted into eq.~(\ref{core9})
the distribution of the core $f$-value sum into the $^3S^o$, $^3P^o$ 
and $^3D^o$ manifolds is given in the proportion 
$\frac{1}{9}:\frac{3}{9}:\frac{5}{9}$.  This is of
course just the statistical weighting associated with
the $(2L_T+1)$ degeneracy factor.  It is simple
to verify that such a proportion means the net contribution
of the core to the tensor polarizability as defined 
by eq.~({\ref{alpha2L}) is zero.

\subsection{The van der Waals coefficients} 

The van der Waals coefficients given in Table \ref{properties}  
are those for a pair of $^3P^o_0$ states.  The dispersion 
parameter, $C_6$ is simple to compute since both of the atoms 
have a net angular momentum of zero. The expression is 
\begin{equation}
C_6  =  \frac{3}{2} \sum_{n_1,n_2}    \frac {f_{0,n_1} f_{0,n_2} } 
         {\epsilon_{0n_1} \epsilon_{0n_2} (\epsilon_{0n_1}+\epsilon_{0n_2})}  \ .  
\label{C6} 
\end{equation}
The present dispersion parameters are slightly larger than 
those of the SG-CI calculation of Santra and Greene  
\cite{santra04a}.  Taking the case of Sr, the difference here
is about 15$\%$.  About half of this difference can be 
attributed to the core since $C_6 = 5668$ au when core 
excitations are omitted from eq.(\ref{C6}). So part of the 
discrepancy arises from the neglect of the core in the SG-CI  
calculation.   One cautionary note should be made.  Santra 
and Greene reported $C_6$ for the $^3P^o_0$ state.   
Since the $^3P^o_0$ state is the most tightly bound state of 
the $5s5p$ multiplet one expects the present $LS$ coupled 
calculation to have a slightly
larger $C_6$.   The quantitative impact of spin-orbit
splitting can best be determined by separate evaluations of 
$C_6$ for the $J = 0, 1$ and $2$ states. 

The van der Waals coefficients that are relevant to BEC 
studies are those between two $^3P^o_2$ states.  The algebra 
related to this is somewhat messy and the coefficients
are presented in the formalism of Santra and Greene 
\cite{santra03a,santra04a}.  
The intermediate dispersion coefficient between two $^3P^o_{J}$ 
states is defined as    
\begin{equation}
B_{J_1,J_2}  = (-1)^{J_1-J_2+1}  
               \sum_{n_1,n_2} \frac {f_{0,n_1} f_{0,n_2} } 
               {\epsilon_{0n_1} \epsilon_{0n_2} (\epsilon_{0n_1}+\epsilon_{0n_2})} \ ,   
\label{BJJsum} 
\end{equation}
where $n_1$ has angular momentum $J_1$ and $n_2$ has angular momentum $J_2$.

This $LS$ coupled oscillator strengths were converted into the $LSJ$ 
coupling scheme using the identity  
\begin{equation}
f(J_0 \to J_{n} ) = f(L_0 \to L_n) (2L_0+1)(2J_n+1)  
\begin{Bmatrix} S & L_0 & J_0 \\ 1 & J_n & L_n \end{Bmatrix}^2  \ .  
\label{LStoJJ} 
\end{equation}
When the sum, eq.~({\ref{BJJsum}), was evaluated, the core $f$-value 
distribution was included using eq.~(\ref{core9}) to partition it into 
$^3S^e$, $^3P^e$ and $^3D^e$ excitations.  

The results of our calculations are presented in Table \ref{B6} and
compared with earlier CI+MBPT calculations of Derevianko {\em et al} 
\cite{derevianko03a} and the SG-CI calculations 
\cite{santra04a}.  There is no apparent experimental activity on the
metastable states of Be and the present data in the Table were only 
included for reasons of completeness.

The present calculation and the CI+MBPT calculation could hardly
be in any better agreement for magnesium.  The largest disagreement
for any of the $B_{J_1,J_2}$ coefficients was $1.2\%$ for the 
$B_{2,2}$ coefficient.  Agreement with the SG-CI calculation
is not as good with the occasional discrepancy of 5$\%$ and it
is noticeable that the present and CI+MBPT results do tend
to be larger in magnitude.     
   
For calcium there is a tendency for the present results to be 
from $1\%$ to $5\%$ larger in magnitude with the differences being
smaller for the larger values of $J_1$ and $J_2$.  The present
dispersion coefficients all lie within the 10$\%$ uncertainty 
that Derevianko {\em et al} associate with their results.  
Agreement with the SG-CI calculations is not so good with
discrepancies exceeding 10$\%$ being common.

The pattern for strontium is similar to that seen for calcium.
The present $B_{J_1,J_2}$ coefficients are larger than 
the CI+MBPT data for $B_{1,1}$ and smaller for $B_{3,3}$.       
The differences with the SG-CI calculation are generally 
larger than those with the CI+MBPT calculation.    

Some general trends are noticeable.  The SG-CI calculation
always gave the smallest result for $B_{1,1}$, $B_{2,1}$, 
$B_{3,1}$, $B_{3,2}$ and $B_{3,3}$.  Furthermore, the sum 
$\sum_{J_i,J_j} B_{J_i,J_j}$ for the present calculations  
and CI+MBPT calculations are consistently bigger than the 
SG-CI calculations, with the difference becoming larger as
the atom gets heavier. This could be a manifestation
of the increasing importance of the core contribution to 
the $B_{J_i,J_j}$ coefficients as the atom gets heavier.   

It is also evident that some of the uncertainty estimates 
of the SG-CI calculation were somewhat optimistic.  For
example, they give $B_{1,1} = 139\pm7$ au for strontium.  
The contribution of the core $f$-value sum to this dispersion 
parameter is 12.0 au.  So the core contribution, which 
is not incorporated in the SG-CI calculation, is  
larger than their estimated uncertainty.

\begingroup
\squeezetable 
\begin{table*}[th]
\caption[]{  \label{B6}
The intermediate dispersion coefficients, $B_{J_1,J_2}$ for  
two alkaline-earth-metal atoms in the metastable $^3P^o_2$ state.    
The $\sum |B_{J_i,J_j}|$ column sums the absolute value of
all the entries in each row (with off-diagonal elements 
added twice).  The numbers in brackets after 
the data are the uncertainties ascribed to the CI+MBPT and 
SG-CI calculations.  
}
\vspace{0.1cm}
\begin{ruledtabular}
\begin{tabular}{lccccccc}
Method & $B_{1,1}$ &  $B_{2,1}$ & $B_{2,2}$ &  $B_{3,1}$ &  $B_{3,2}$ & $B_{3,3}$ & $\sum |B_{J_i,J_j}|$ \\ \hline
\multicolumn{8}{c}{Be} \\ 
Present  & -6.901 & 11.29 & -27.77  & -13.52  & 22.31  & -27.21  & 220.3 \\ \hline 
\multicolumn{8}{c}{Mg} \\ 
Present  & -37.46 & 43.90 & -52.73  & -76.49  & 90.76  & -157.2  & 1004 \\
CI+MBPT \cite{derevianko03a} & -37.8(38)  & 43.9(44)   & -52.1(52) & -76.7(77)  & 90.1(90)  & -156.4(156) & 1002  \\  
SG-CI \cite{santra04a} &  -35.6(2) & 42.5(2)   & -51.9(2)  & -73.4(7)  &  88.6(7)  & -152(2) & 976 \\ \hline  
\multicolumn{8}{c}{Ca} \\ 
Present  & -96.95 & 130.7 & -176.8  & -233.5  & 317.5  & -604.8 & 3363  \\
CI+MBPT \cite{derevianko03a} & -91.7(92)  & 123(12)   & -167(17) & -225(23)  & 306(31)  & -600(60) & 3250 \\  
SG-CI \cite{santra04a} &  -81(3) &   119(5)   & -176(8)  & -203(10)  &  302(20)  & -553(70) & 3087 \\ \hline  
\multicolumn{8}{c}{Sr} \\ 
Present  & -165.9 & 213.9   & -278.9 &  -416.9 & 556.0 & -1231 & 6074 \\
CI+MBPT \cite{derevianko03a} & -158(16)  & 203(20)   & -264(26) & -415(42)  & 555(56) & -1290(130) & 6090 \\  
SG-CI \cite{santra04a} &  -139(7) &   196(9)   & -280(10)  & -370(30)  &  546(50)  & -1210(200)& 5780 \\  
\end{tabular} 
\end{ruledtabular}
\end{table*}
\endgroup

\section{Conclusions}

A systematic study of the properties of the alkaline-earth atoms reveals
that the present non-relativistic approach reproduces the results of the 
CI+MBPT ansatz of Derevianko {\em et al} \cite{derevianko03a} to better 
than 5$\%$.  Agreement with the model potential CI calculation
of Santra and Greene \cite{santra04a} is not so good with discrepancies
of 10-15$\%$ occuring for the spherical part of the $C_6$ dispersion 
coefficient.   Due to the unknown impact of the spin-orbit energy,
splitting upon the polarizabilities and dispersion coefficients, it 
is not possible to make a definitive statement about any reasons 
for the differing levels of agreement with these two other calculations.  
However, we do suspect that the omissions of the SG-CI model, 
i.e. the di-electronic two body polarization potential, the non-usage 
of a dressed dipole transition operator, and the lack of core excitation
terms in the dispersion sum rules all contribute in part to the 
differences with the SG-CI model.    

It should be noted that previous studies with the present model for the 
alkali atoms and singlet states of the alkaline atoms demonstrated
that the method could predict a number of expectation values with an 
overall accuracy of 1-2$\%$ or better \cite{mitroy03f}.  The
presence of spin-orbit energy splitting, and the existence of 
a $^3D^e$ state very close in energy to the $^3P^o$ metastable
level leads to a decrease in accuracy for atomic properties 
such as the tensor polarizability that are sensitive to these 
energy differences. Additional high precision measurements of 
the tensor polarizabilities for the Mg, Ca and Sr would be certainly 
be worthwhile since the sensitivity of this parameter to the fine 
details of the wave function should help in the refinement of the 
two-body potentials used to characterize ultra-cold collisions.       

It is interesting to speculate whether the better agreement
with the CI+MBPT calculations could be achieved by incorporating
a spin-orbit potential into the Hamiltonian and using $jj$ 
coupling.  Alternatively, a fully relativistic treatment, 
using a relativistic HF wave function might be necessary.  
Resolution of these questions requires that explicit calculations 
be made to determine the additional physics needed to eliminate 
the anomalies between the present calculations and experiment and 
between the present calculations and the CI+MBPT calculations.

\section{Acknowledgments}

The authors would like to thank Mr J C Nou and Mr C Hoffman  
of CDU for workstation support.


\end{document}